\documentclass[12pt]{article}

 \usepackage{amsmath}
 \usepackage{graphicx}
 \usepackage{amssymb}
 \usepackage{amsfonts}
 \usepackage{latexsym}

\renewcommand{\baselinestretch}{1.2}
\def\beq{\begin{eqnarray}}
\def\eeq{\end{eqnarray}}

\def\Ln{\,\mbox{Ln}\,}
\def\Det{\,\mbox{Det}\,}

\def\tr{\,\mbox{tr}\,}

\def\Tr{\,\mbox{Tr}\,}

\def\al{\alpha}
\def\be{\beta}
\def\ch{\chi}

\def\de{\delta}

\def\la{\lambda}
\def\na{\nabla}

\def\rh{\rho}
\def\si{\sigma}

\def\ph{\varphi}
\def\ta{\tau}
\def\th{\theta}

\def\Ga{\Gamma}

\def\La{\Lambda}

\begin{document}

\begin{center}

{\large{\bf One-loop divergences in massive gravity theory}}

\vskip 6mm


\textbf{I. L. Buchbinder,}$^{\,\dagger,\ddagger,}
$\footnote{
E-mail: joseph@tspu.edu.ru}
\quad
\textbf{D. D. Pereira}$^{\,\ddagger,}
$\footnote{
E-mail: dante.pereira@fisica.ufjf.br}
\quad
\textbf{and}
\quad
\textbf{I. L. Shapiro}$^{\,\ddagger,}
$\footnote{
E-mail: shapiro@fisica.ufjf.br. On leave from Tomsk State
Pedagogical University, Tomsk, Russia.}
\vskip 6mm
$^{\,\dagger}$ Department of Theoretical Physics,
Tomsk State Pedagogical University, 634041, Tomsk, Russia
\vskip 4mm
$^{\,\ddagger}$ Departamento de F\'{\i}sica, ICE,
Universidade Federal de Juiz de Fora, 36036-330, MG, Brazil
\end{center}

\vskip 12mm

\begin{quotation}
\noindent {\large {\it Abstract}}.
\quad
The one-loop divergences are calculated for the recently proposed
ghost-free version of massive gravity, where the action depends
on both metric and external tensor field $f$. The non-polynomial
structure of the massive term is reduced to a more standard form
by means of auxiliary tensor field, which is settled on-shell
after quantum calculations are performed. As one should expect,
the counterterms do not reproduce the form of the classical action.
Moreover, the result has the form of the power series in $f$.
\end{quotation}
\vskip 4mm

\section{Introduction}
\label{int} It is well known that general relativity is a non-linear
dynamical theory of symmetric second rank tensor field in curved
space-time. In the linear approximation this theory describes a
propagation of massless spin-2 irreducible representation of the
Poincare group. Dynamical theory of massive spin-2 representation of
the Poincare group has been constructed by Fierz and Pauli \cite{FP}
in terms of symmetric second rank tensor field in Minkowski space.
It is natural to think that there should exist a non-linear dynamical
theory in term of symmetric second rank tensor field in curved
space-time, whose linear limit will be the Fierz-Pauli theory.
However, during a long time such theory has not been constructed.

An evident way to find non-linear generalization of Fierz-Pauli
theory is to add some kind of massive term into Lagrangian of
general relativity. It could be a cosmological constant, but in
the linear limit the cosmological term does not provide a true mass
term in Fierz-Pauli theory. As a result the only way to insert a
mass term to Lagrangian of general relativity is to use, additionally
to metric, an extra second rank tensor field (reference metric) and
find a coupling of metric to this extra field in such a way that in
the linear approximation for both metric and Minkowski reference
metric, this coupling should generate a true mass term in Fierz-Pauli
theory.

The extension of general relativity described above has been studied
in details by Boulware and Deser \cite{BD}. They have shown that
inserting the mass term with the help of additional second rank
tensor field can yield, in general, the inconsistent theory with
propagating ghost field (BD ghost). However recently there was a
progress in constructing a family of the non-linear dynamical
theories which are free of problem of BD ghost \cite{ghost free-1},
\cite{ghost free-2}, \cite{ghost free-3}, \cite{ghost free-4} (see
also the review \cite{review}). As a result, at present one can have
the Lagrangians which satisfy the following set of conditions:
(i) describe consistent ghost-free non-linear dynamics of symmetric
second rank tensor field; \ \
(ii) have a number of propagating degrees of freedom exactly
corresponding to massive spin-2 field, \ \ (iii) depend of reference
metric and on some mass parameter $m$, while in the limit $m=0$
reproduce the Lagrangian of general relativity without cosmological
term; \ \ (iv) reproduce the Lagrangian of Fierz-Pauli theory in
linear approximation for dynamical metric and for Minkowski
reference metric.

In this note we study the quantum aspects of the massive gravity
theory. To be more precise, we compute the one loop divergences of a
minimal massive theory \cite{ghost free-2} and investigate their
structure. Of course, the massive gravity theory is
non-renormalizable as well as general relativity since massive
gravity Lagrangian includes general relativity Lagrangian and
hence should leat to the quadratic in curvature tensor counter-terms
at one loop. In order to find the one-loop divergences we will use
the background field method and Schwinger-DeWitt proper-time
techniques \cite{DW-65}.

Although the theory under consideration is non-renormalizable, one
can expect it to have many interesting features in quantum domain.
First, the massive term in the action possesses a complicated
structure in tensor indices and its background-quantum splitting
(decomposition of initial field into background and quantum field,
which is the element of the background field method) is not trivial.
Second, a functional determinant, defining the one-loop effective
action, has different and in fact more complicated structure as
compared to the massless case of general relativity. Therefore,
evaluation of effective method may require some novelty in the
method of calculation. Third, it is interesting to study whether
the one-loop divergences in massive gravity theory vanish on-shell
as it is the case for the general relativity without matter. Fourth,
the one-loop divergences in massive gravity theory are expected to
depend on the reference metric. As a result of computations we obtain
the one-loop divergences in terms of the specific general covariant
functionals depending on reference metric. They can be considered as
the possible candidates for actions of reference metric if to treat
this metric as a dynamical field. Problem of Lagrangian for
reference metric is broadly discussed in the literature (see e.g.
\cite{ghost free-2} and reference therein). Fifth, any new
gravitational model deserves a study of quantum aspects in the
hope to extend our understanding of a quantum gravity and to
shead a light on a possible role of gravity in quantum domain.

The paper is organized as
follows. In Sect. 2 we consider an equivalent representation of the
theory via auxiliary tensor field and the procedure of
linearization. The derivation of bilinear form of the action and the
details of background field method are treated in Sect. 3. In Sect.
4 we present the results of calculating the one-loop divergences. In
Sect. 5 we present some discussions of the result and draw our
conclusions.

\section{Linearization of massive term}
\label{massive gravity}

Consider the action of a minimal model of massive gravity
\cite{ghost free-2}
\beq \nonumber
S[g_{\mu\nu}]&=&S_{0}[g_{\mu\nu}]+S_{m}[g_{\mu\nu}]
\\
&=&
\int d^{4}x \sqrt{- g}\, \Big[R + 2\La \,+\,m^{2}\tr
\left(\sqrt{g^{-1}f}\right)\Big]\,.
\label{1}
\eeq
Here $m$ is a
mass parameter, $f$ means an external tensor field (reference
metric) $f_{\mu \nu}$, expression $g^{-1}f$ in action $S_{m}$ means
$g^{\mu\alpha}f_{\alpha\nu}$, all other notations are
standard\footnote{The gravitational constant ${\kappa}$ is
suppressed}.
According to \cite{ghost free-2} we have to put $\La=-3m^2$, however
it is more convenient to perform all calculations for an arbitrary
value of $\La$ and fix it only in the final result.

To calculate the one-loop divergences of the theory under
consideration we should compute the second variational derivative of
action with respect $g_{\mu\nu}$. Such computation for the term
$S_{m}$ is very non-trivial since the matrices $g^{\mu\alpha}$ and
$f_{\alpha\nu}$ do not commute\footnote{The first variational
derivative is computed simply enough \cite{ghost free-2}, however
the second derivative can not be computed analogously to \cite{ghost
free-2}.}. We will avoid this obstacle considering the classically
equivalent theory formulated in terms of dynamical metric
$g_{\mu\nu}$ and auxiliary field $\ph^\mu{}_\nu$.

It is easy to see that the theory, described by the action (\ref{1})
in terms of dynamical field $g_{\mu\nu}$, is classically equivalent
to the theory in terms of the fields $g_{\mu\nu}$ and auxiliary
field $\ph^\mu{}_\nu$ with the following action
\beq
\widetilde{S}[g_{\mu\nu},\varphi^{\mu}{}_{\nu}]
&=&  S_{0}[g]+\widetilde{S}_{m}[g_{\mu\nu},\varphi^{\mu}{}_{\nu}],
\label{2}
\eeq
where the action
$\widetilde{S}_{m}[g_{\mu\nu},\varphi^{\mu}{}_{\nu}]$ is given by
\beq
\widetilde{S}_{m}[g_{\mu\nu},\varphi^{\mu}{}_{\nu}]
\,=\,\frac{m^{2}}{2}\int d^{4}x \sqrt{- g}\,
\big[ g^{\mu \nu}f_{\nu\al}
\varphi^{\al}{}_{\mu} +(\varphi^{-1})^{\mu}{}_{\mu} \big].
\label{3}
\eeq

Indeed, the equation of motion that follows from the variation over
$\varphi^{\mu}{}_{\nu}$ in (\ref{2}) has the form
\beq
\frac{\de \widetilde{S}}{\de \varphi^{\mu}{}_{\nu}} =\frac{\de
\widetilde{S}_{m}}{\de \varphi^{\mu}{}_{\nu}}=0 \,, \quad
\nonumber
\eeq
The solution to this equation is
\beq
\quad \ph^{\mu}{}_{\nu} =
\big[\left(g^{-1}f\right)^{-1/2}\big]^{\mu}{}_{\nu}.
\nonumber
\eeq
Replacing this solution back into (\ref{2}), we obtain the
action (\ref{1}). It shows that the two actions are classically
equivalent. Starting from this point we will use the action
(\ref{2})\footnote{We assume that this equivalence is fulfilled on
the quantum level as well, since there is no any source for possible
anomaly}.

\section{Background field method and bilinear form to action}
\label{bfm}

Our main purpose is to develop the background field method (
\cite{DW-65}, see also the details in \cite{article 9}) to the
theory (\ref{2}) and use it to calculate the one-loop divergences of
the theory. The first step is to obtain the bilinear form of the
action (\ref{2}).

According to the background field method, the fields
$\ph^{\mu}{}_{\nu}$ and $g_{\mu \nu}$ are replaced by sums of
background and quantum fields as follows
\beq
\ph^{\mu}{}_{\nu}\rightarrow \ph^{\mu}{}_{\nu}+\psi^{\mu}{}_{\nu},
\qquad g_{\mu \nu}\rightarrow g_{\mu \nu}+h_{\mu \nu}.
\label{4}
\eeq
Here $\ph^{\mu}{}_{\nu}$ and $g_{\mu \nu}$ are background
fields, while $\psi^{\mu}{}_{\nu}$ and $h_{\mu \nu}$ are quantum
fields. By means of a simple algebra one can obtain the bilinear
form for the actions ${S}_{0}$ and $\widetilde{S}_{m}$ in the form
\beq
{S}_{0}^{(2)} + S_{gf} \,=\, \frac{1}{2}\,
\int d^{4}x \sqrt{-g}\, h^{\mu \nu}
\,\hat{J}_{\mu \nu ,\al \be}\,h^{\al \be},
\label{5} \eeq and \beq {\widetilde{S}}_{m}{}^{(2)} \,=\,
m^{2}\,\int d^{4}x \sqrt{- g}\Big\{ \, \frac{1}{2}h^{\al
\be}\,\hat{G}_{\al \be ,\mu \nu}\,h^{\mu \nu}
-\frac{1}{2}\psi^{\al}{}_{\be}\,\hat{A}_{\al}{}^{\be},_{\mu}{}^{\nu}\,
\psi^{\mu}{}_{\nu}
\,+\,\hat{B}^{\be}{}_{\al}\,\psi^{\al}{}_{\be}\Big\}.
\label{6}
\eeq
In Eq. (\ref{5}) we have introduced the so-called minimal gauge
fixing term \beq S_{gf}=-\frac{1}{2}\int d^4x \sqrt{- g}\;
\ch^{\mu}\ch_{\mu}, \eeq
 where
 \beq
\ch_{\mu}=\nabla_{\la}h^{\la}{}_{\mu}-\frac{1}{2}\nabla_{\mu}h\,.
\label{7} \eeq

The operators \ $\hat{J}_{\mu \nu ,\al \be}$, \
$\hat{G}_{\al \be ,\mu \nu}$, \
$\hat{A}_{\al}{}^{\be},_{\mu}{}^{\nu}$ \ and \
$\hat{B}^{\be}{}_{\al}$, which were used in (\ref{5}) and
(\ref{6}), have the form
\beq
\nonumber
\hat{J}_{\mu \nu ,\al \be}
&=& \frac{1}{2}K_{\mu \nu ,\al \be}\Box
+ R_{\mu \al \nu \be} +g_{\nu \be}R_{\mu \al}
- \frac{1}{2}(g_{\mu \nu}R_{\al \be}
+ g_{\al \be}R_{\mu \nu})
- \frac{1}{2} (R+2\La) K_{\mu \nu ,\al \be},
\\
\nonumber
\hat{G}_{\al \be ,\mu \nu}
&=& -\frac{1}{4}K_{\al \be ,\mu \nu}
\Big[ g^{\si \ta}f_{\ta \la}\varphi^{\la}{}_{\si}
+(\varphi^{-1})^{\si}{}_{\si}\Big]
- \frac{1}{2}g_{\al \be}f_{\nu \si}\varphi^{\si}{}_{\mu}
+g_{\mu \be}f_{\nu \si}\varphi^{\si}{}_{\al},
\\
\nonumber
\hat{A}_{\al}{}^{\be},_{\mu}{}^{\nu}
&=&
- \frac{1}{2}
\Big[
(\varphi^{-1})^{\si}{}_{\al}(\varphi^{-1})^{\be}{}_{\mu}
(\varphi^{-1})^{\nu}{}_{\si}
+(\varphi^{-1})^{\si}{}_{\mu}(\varphi^{-1})^{\nu}{}_{\al}
(\varphi^{-1})^{\be}{}_{\si}\Big],
\\
\hat{B}^{\be}{}_{\al}&=&-\frac{1}{2}\,h^{\be \la}f_{\la \al},
\label{8}
\eeq
where we define
\beq
K_{\al \be ,\mu \nu} = \de_{\al\be ,\mu \nu}
- \frac{1}{2}g_{\al \be}g_{\mu \nu}
\label{9}
\eeq
and use notation
\beq
\de_{\al \be ,\mu \nu}
= \frac{1}{2}(g_{\al\mu}g_{\be \nu}
+ g_{\al \nu}g_{\be \mu}) \nonumber
\eeq
for the DeWitt identity matrix in the space of symmetric matrices.
In the expressions (\ref{8}) one have to assume symmetrization in both
couples of indices \ $\mu\nu$ \ and \ $\al\be$. Let us note that the
expression for \ $\hat{J}_{\mu \nu ,\al \be}$ \ in (\ref{8}) is one
for the usual Einstein quantum gravity with the cosmological
constant and the other terms here are because of the massive
terms in Eq. (\ref{1}).

It is easy to see that the path integral over the quantum
field $\psi^{\mu}{}_{\nu}$ can be taken at once. It is well
known that the following identity holds for Hermitian
matrices $A(y,x)$:
\beq
\nonumber
\int {\cal{D}}\psi
\exp \Big\{
-\frac{i}{2}\int dy\int dx
\,\psi(y)A(y,x)\psi(x)+ i \int dx \,B(x)\psi(x)
\Big\}
\\
= \big(\Det A\big)^{-1/2}\,\times\,
\exp\Big\{ \frac{i}{2}\int dy\int dx\, B(y)\,A^{-1}(y,x)\,B(x)\Big\}.
\label{10}
\eeq
Let us
note that the quantity $\big(\Det A\big)^{-1/2}$ corresponds to the
determinant of a numerical matrix. Since we assume dimensional
regularization here, this object is irrelevant to the analysis of
quantum corrections to the effective action and therefore will not
be omitted. Using Eq. (\ref{10}) in the expression for the
generating functional of Green functions, we present the bilinear
form for the action (\ref{2}) as follows
\beq
\nonumber
\widetilde{S}{}^{(2)} &=& S_{0}^{(2)} \,+\, S_{gf} \,+\,
\widetilde{S}_{m}^{(2)} = \frac{1}{2} \int d^4x \sqrt{- g}\,\,
h^{\al \be}\,\hat{H}_{\al \be , \mu \nu}\,h^{\mu \nu}\,,
\label{11}
\eeq
where the operator $\hat{H}_{\al \be , \mu \nu}$ is given by
\beq
\hat{H}_{\al \be , \mu \nu}=\hat{J}_{\mu \nu ,\al \be}
+m^{2}\hat{G}_{\al \be ,\mu \nu} +\frac{1}{4}m^2 f_{\be \la}
(\hat{A}^{-1})_{\al}{}^{\la},_{\mu}{}^{\si}f_{\nu \si}\,.
\label{12}
\eeq

In order to obtain the matrix $A^{-1}$, let us consider the
following procedure. The result (\ref{10}) is valid for the
Hermitian matrix $A$. Therefore we need to take only symmetric
part of the matrix $A$. Consider first the matrix ${\bar A}$
which is not symmetrized. It is an easy exercise to find its
inverse, however one has to work a little bit more to do the
same with the symmetric part of it. One can write ${\bar A}$ as
\beq
\bar{A}_{\al}{}^{\be},_{\mu}{}^{\nu}=
-(\varphi^{-1})^{\si}{}_{\al}(
\varphi^{-1})^{\be}{}_{\mu}(\varphi^{-1})^{\nu}{}_{\si}.
\label{13-1}
\eeq
The corresponding inverse matrix is given by
\beq
{({\bar{A}}^{-1})}{}_{\be}{}^{\al},_{\si}{}^{\rh}
\,=\, -
{\varphi}^{\rh}{}_{\tau}{\varphi}^{\tau}{}_{\be}{\varphi}^{\al}{}_{\si}.
\label{13-2}
\eeq
Consider now the following symmetric structure
\beq
{X}{}_{\be}{}^{\al},_{\si}{}^{\rho}=
-{\varphi}^{\rho}{}_{\tau}
{\varphi}^{\tau}{}_{\be}{\varphi}^{\al}{}_{\si}
-{\varphi}^{\al}{}_{\tau}
{\varphi}^{\tau}{}_{\si}{\varphi}^{\rho}{}_{\be}.
\label{14-1}
\eeq
Then we arrive at the equation
\beq
\hat{A}_{\mu}{}^{\nu},_{\al}{}^{\be}
\times {X}{}_{\be}{}^{\al},_{\si}{}^{\rho}
\,=\, {Z}{}_{\mu}{}^{\rho},_{\si}{}^{\nu}
=\de^{\rho}_{\mu}\de^{\nu}_{\si}+{Y}{}_{\mu}{}^{\rho},_{\si}{}^{\nu},
\label{14-2}
\eeq
where
\beq
{Y}{}_{\mu}{}^{\rho},_{\si}{}^{\nu}
\,=\,
\frac{1}{2}{({\varphi}^{-1})}^{\rho}{}_{\mu}{\varphi}^{\nu}{}_{\si}
+ \frac{1}{2}{({\varphi}^{-1})}^{\nu}{}_{\si}{\varphi}^{\rho}{}_{\mu}.
\label{15}
\eeq

Finally, we have the inverse to the symmetrized matrix in the
form of the series
\beq
{({\hat{A}}^{-1})}_{\mu}{}^{\nu},_{\al}{}^{\be}=
{X}{}_{\mu}{}^{\nu},_{\la}{}^{\si}\,\times \,
(Z^{-1}){}_{\si}{}^{\la},_{\al}{}^{\be},
\label{16}
\eeq
where the matrix
$(Z^{-1}){}_{\si}{}^{\la},_{\al}{}^{\be}$ is given by
\beq
(Z^{-1}){}_\si {}^\la ,_\al{}^\be
\,=\,
\de^{\la}_{\si}\de^{\be}_{\al}
\,-\,{Y}{}_{\si}{}^{\la},_{\al}{}^{\be}
\,+\,{Y}{}_{\si}{}^{\la},_{\rho}{}^{\tau}\,
{Y}{}_{\tau}{}^{\rho},_{\al}{}^{\be}
\,-\, {Y}{}_{\si}{}^{\la},_{\rho}{}^{\tau}\,
{Y}{}_{\tau}{}^{\rho},_{\ch}{}^{\de}\,
{Y}{}_{\de}{}^{\ch},_{\al}{}^{\be} + \,...\,\,.
\label{17}
\eeq
Multiplying the operator $\hat{H}_{\al \be , \mu \nu}$ by the
operator \ $2\hat{K}^{-1}{}_{\la \si},^{\al \be}$, \ where
\beq
\hat{K}^{-1}{}_{\la \si},^{\al \be}=
{\de}_{\la \si},^{\al \be}
-\frac{1}{2}g_{\la \si}g^{\al \be},
\nonumber
\eeq
we arrive at
\beq
2 \hat{K}^{-1}{}_{\al \be},^{\la \si}
{\hat{H}}_{\la \si ,\mu \nu}\equiv
{\hat{O}}_{\al \be ,\mu \nu}=
\de_{\al \be , \mu \nu}\Box
+\hat{\Pi}_{\al \be , \mu \nu},
\label{18}
\eeq
where we have

\beq
\nonumber
\hat{\Pi}_{\al \be , \mu \nu}&=&
2R_{\al \mu \be \nu}
+2g_{\be \nu}R_{\al \mu}
-g_{\al \be}R_{\mu \nu}
-g_{\mu \nu}R_{\al \be}
-R K_{\al \be, \mu \nu}
-2\La \de_{\al \be, \mu \nu}\\ \nonumber
&+&
\frac{m^{2}}{2}f_{\rh (\al}
(\hat{A}^{-1})_{\be )}{}^{\rh},_{\mu}{}^{\tau}
f_{\nu \tau}
-
\frac{m^{2}}{4}g_{\al \be}f_{\si \rh}
(\hat{A}^{-1})^{\si \rh},_{\mu}{}^{\tau}
f_{\nu \tau}
+2 m^{2}f_{\nu \si}\varphi^{\si}{}_{(\be}g_{\al )\mu}
\\
&-&\frac{m^{2}}{2}\de_{\al \be ,\mu \nu}
\Big[ g^{\si \ta}f_{\ta \la}\varphi^{\la}{}_{\si}
+(\varphi^{-1})^{\si}{}_{\si}\Big].
\label{19}
\eeq
\section{Derivation of one-loop divergences}
\label{corrections}

The one-loop quantum corrections to effective action is written by
standard way (see e.g. \cite{article 6})
\beq
\overline{\Ga}^{(1)}
&=& \frac{i}{2}\,\Ln \Det(\hat{H})
\,=\,\frac{i}{2}\,\Tr \Ln(\hat{H}),
\label{20}
\eeq
where the operator $\hat{H}$
corresponds to the bilinear part of the action in quantum fields and
$\Tr$ means the functional trace.

The divergent part of $\Tr \Ln \hat{H}$ can be obtained
by calculating $\Tr \Ln (\hat{K}^{-1}\hat{H})$ and then
subtracting the $\Tr \Ln \hat{K}^{-1}$
\beq
-\Tr \Ln \hat{H}=
-\Tr \Ln (\hat{K}^{-1}\hat{H})
+\Tr \Ln \hat{K}^{-1}.
\label{21}
\eeq
However, as far as we are interested in the logarithmic
divergent part of the effective action, the contribution
of the last term can be safely omitted.

The computation of (\ref{20}) can be performed by the use of the
Schwinger-DeWitt proper-time technique \cite{DW-65}. This technique
provides the efficient method of evaluating the $\Ln \Det \hat{O}$,
where the operator $\hat{O}$ has the form (see e.g. \cite{hove},
\cite{BV})
\beq
\hat{O} = \hat{1}\Box
+ 2\hat{h}^{\mu}\na_{\mu} + \hat{\Pi}.
\label{22}
\eeq
Also we
introduce the operators
\beq
\hat{P} & = & \hat{\Pi}
+\frac{1}{6}R\hat{1} -\na_{\mu}\hat{h}^{\mu}
-\hat{h}^{\mu}\hat{h}_{\mu}, \nonumber
\\
\hat{S}_{\mu\nu} & = & \,=\, \hat{1}[\na_{\mu},\na_{\nu}] +
\na_{\nu}\hat{h}_{\mu} - \na_{\mu}\hat{h}_{\nu}+
[\hat{h}_{\nu},\hat{h}_{\mu}].
\label{23}
\eeq
In our case, exactly
as for the Einstein quantum gravity, the $\hat{h}_{\mu}=0$ and this
essentially simplifies the calculations.

In the framework of dimensional regularization, the quantity
${\bar \Ga}^{(1)}$
is written as follows
\beq
{\bar \Ga}^{(1)}_{div} \,=\,
-\frac{\mu^{n-4}}{(4\pi)^{2}(n-4)}\Tr\left\{\frac{\hat{1}}{180}
(R^2_{\mu\nu\al\be} - R^2_{\mu\nu}) + \frac{1}{2}\hat{P}^2 +
\frac{1}{12} {\hat{S}^2}_{\mu\nu} \right\},
\label{24}
\eeq
where
$\mu$ is the parameter of dimensional regularization, the operators
$\hat{P}$, $\hat{S}_{\mu\nu}$ are defined above and the surface
terms are ignored.

In our case the operators $\hat{P}$ and $\hat{S}_{\mu\nu}$
have the form
\beq
\nonumber
\hat{P}_{\al \be , \mu \nu}&=&
2R_{\al \mu \be \nu}
+2g_{\be \nu}R_{\al \mu}
-g_{\al \be}R_{\mu \nu}
-g_{\mu \nu}R_{\al \be}
-\frac{5}{6}R \de_{\al \be, \mu \nu}
\\
\nonumber
&+& \frac{1}{2}Rg_{\al \be}g_{\mu \nu}
-2\La \de_{\al \be, \mu \nu}
-\frac{m^{2}}{2}\de_{\al \be ,\mu \nu}
\Big[ g^{\si \ta}f_{\ta \la}\varphi^{\la}{}_{\si}
+(\varphi^{-1})^{\si}{}_{\si}\Big]
\\
\nonumber
&+&
2 m^{2}f_{\nu \si}\varphi^{\si}{}_{(\al}g_{\be )\mu}
+ \frac{m^{2}}{2}f_{\rh (\al}
(\hat{A}^{-1})_{\be )}{}^{\rh},_{\mu}{}^{\tau} f_{\nu \tau}
- \frac{m^{2}}{4}g_{\al \be}f_{\si \rh}
(\hat{A}^{-1})^{\si \rh},_{\mu}{}^{\tau}
f_{\nu \tau},
\\
\hat{S}_{\la \tau} &=& {[\hat{S}_{\la \tau}]}_{\mu \nu ,\al \be}
\,=\, -\,2\,R_{\mu \al \la \tau}g_{\nu \be}\,.
\label{25}
\eeq
Replacing these operators in the expression (\ref{24}), after some
algebra we obtain the expression for the divergent part of the
one-loop effective action,
\beq
\nonumber
&&
\overline{\Ga}^{(1)}_{div}\mid_{\La=-3m^2}
\,=\,-\,\frac{2\mu^{n-4}}{(4\pi)^{2}(n-4)} \int d^{n}x \sqrt{- g}
\Big\{\frac{53}{90}E +\frac{7}{20}R_{\mu \nu}^{2}
+\frac{1}{120}R^{2}
\\
\nonumber
&+& \frac{m^2}{2}R^{(\la \mid\al\mid \si ) \be}f_{\si \tau}
{({\hat{A}}^{-1})}{}_{\la}{}^{\tau}{}_{(\al}{}^{\mid\rh\mid}
f_{\be )\rh}
- \frac{m^2}{48}R\Big[948 + 236\big(\sqrt{g^{-1}f}\big)^{\mu}{}_{\mu}
\\
\nonumber
&+&
+ 5 \big(g^{-1}f\big)^{\la}{}_{\be}
{({\hat{A}}^{-1})}{}_{\al \la ,}{}^{\al \si}
\big(g^{-1}f\big)^{\be}{}_{\si}
+ 5f_{\al \la} {({\hat{A}}^{-1})}{}^{\be \la , \al \si} f_{\be \si}
- 5f_{\al \be}{({\hat{A}}^{-1})}{}^{\al \be ,\mu \nu}f_{\mu \nu}
\Big]
\\
\nonumber
&+&
\frac{m^2}{4}R^{\la \al}\Big[12\big(\sqrt{g^{-1}f}\big)_{\al \la}
+\big(g^{-1}f\big)^{\be}{}_{\rh}
{({\hat{A}}^{-1})}{}_{\la}{}^{\rh}{}_{(\al}
{}^{\mid \tau \mid}f_{\be )\tau}
+ f_{\la \rh}{({\hat{A}}^{-1})}{}^{\be \rh}{}_{(\al}{}^{\mid \tau \mid}
f_{\be )\tau}
\\
\nonumber
&-& 2f_{\si \tau} {({\hat{A}}^{-1})}{}^{\si \tau}{}_{\al}{}^{\rh}
f_{\la \rh}\Big]
+\frac{m^4}{16}\Big[1440+12\big(g^{-1}f\big)^{\la}{}_{\be}
{({\hat{A}}^{-1})}{}_{\al \la ,}{}^{\al \si}
\big(g^{-1}f\big)^{\be}{}_{\si}
\\
\nonumber
&+& 12f_{\al \la} {({\hat{A}}^{-1})}{}^{\be \la , \al \si} f_{\be \si}
-12f_{\al \be}{({\hat{A}}^{-1})}{}^{\al \be ,\mu \nu}f_{\mu \nu}
-240\big(\sqrt{g^{-1}f}\big)^{\mu}{}_{\mu}
+24\big(g^{-1}f\big)^{\mu}{}_{\mu}
\\
\nonumber
&+&
4\big(\sqrt{g^{-1}f}\big)^{\mu}{}_{\mu}
\big(\sqrt{g^{-1}f}\big)^{\nu}{}_{\nu}
+\big(g^{-1}f\big)^{\la}{}_{(\rh}
{({\hat{A}}^{-1})}{}_{\tau ) \la \mu}{}^{\si}
\big(g^{-1}f\big)^{\nu}{}_{\si}
\big(g^{-1}f\big)^{\rh}{}_{\th}
{({\hat{A}}^{-1})}{}^{\tau \th \mu}{}_{\al}
\big(g^{-1}f\big)^{\al}{}_{\nu}
\\
\nonumber
&+&
\big(g^{-1}f\big)^{\la}{}_{(\rh}
{({\hat{A}}^{-1})}{}_{\tau )\la}{}^{\mu}{}_{\si}
\big(g^{-1}f\big)^{\si}{}_{\nu}
\big(g^{-1}f\big)^{\rh}{}_{\th}
{({\hat{A}}^{-1})}{}^{\tau \th \nu}{}_{\al}
\big(g^{-1}f\big)^{\al}{}_{\mu}
\\
\nonumber
&-&
4\big(\sqrt{g^{-1}f}\big)^{\la}{}_{\la}\big(g^{-1}f\big)^{\be}{}_{\ph}
f_{\th (\al}(\hat{A}^{-1})_{\be )}{}^{\th \al \ph}
+ 2\big(\sqrt{g^{-1}f}\big)^{\la}{}_{\la}
f_{\al \be} {({\hat{A}}^{-1})}{}^{\al \be ,\mu \nu}
f_{\mu \nu}
\\
\nonumber
&-&
4f_{\rh \th}\Big(\sqrt{g^{-1}f}\Big)^{\la}{}_{(\al}f_{\ph )\la}
{({\hat{A}}^{-1})}{}^{\rh \th \al \ph}
+2f_{\al \th}f_{\la \ph}\Big(\sqrt{g^{-1}f}\Big)^{\la}{}_{\rh}
{({\hat{A}}^{-1})}{}^{\rh \th \al \ph}
\\
\nonumber
&+&
2f_{\be \la}\big(g^{-1}f\big)^{\be}{}_{\ph}
\big(\sqrt{g^{-1}f}\big)^{\la}{}_{\th}
{({\hat{A}}^{-1})}{}_{\al}{}^{\th \al \ph}
+2\big(g^{-1}f\big)^{\th}{}_{\la}\big(g^{-1}f\big)^{\la}{}_{\ph}
\big(\sqrt{g^{-1}f}\big)^{\al}{}_{\rh}
{({\hat{A}}^{-1})}{}^{\rh}{}_{\th \al}{}^{\ph}
\\
&+&
2\big(g^{-1}f\big)^{\ta}{}_{\th}f_{\la \ph}
\big(\sqrt{g^{-1}f}\big)^{\al}{}_{\ta}
{({\hat{A}}^{-1})}{}^{\la \th}{}_{\al}{}^{\ph}
\Big]
\Big\}.
\label{28}
\eeq
where we included the mass-independent ghost contribution and
used the special value $\La=-3m^2$. In the Eq. (\ref{28})
\ $E=R_{\mu\nu\al\be}^2-4R_{\mu\nu}^2+R^2$ \
is the integrand of the Gauss-Bonnet topological term and the
expression ${({\hat{A}}^{-1})}{}^{\al\be \mu \nu}$ has been
defined in Eq. (\ref{14-2}). One has to note that the matrix
$({\hat A})^{-1}$ is an infinite power series on the external
field $f$ and hence the divergences (\ref{28}) have essentially
non-polynomial structure in this field too.

Let us note that before the use of the condition  $\La=-3m^2$ the
divergences represent the corresponding expression for Einstein
quantum gravity \cite{hove} with the contribution of the
cosmological term and the rest of the expression is due to
additional mass dependent term in the action. The reason for such a
result is that we performed calculations is the situation when the
diffeomorphism symmetry is unbroken. This means we treat
$f_{\mu\nu}$ as external tensor field which does not violate general
covariance of the theory.

An interesting observation concerning the Eq. (\ref{28}) is that
there is an explicit simple hierarchy of the terms, for example the
ones with higher derivatives do not depend on mass and/or on the
field $f$. At the same time, if we consider the classical action
with the algebraic structures presented in (\ref{28}), we note that
there are no derivatives acting on $f$ there. However, despite there
are no such derivatives of $f$, this field will be dynamical in
action (\ref{28}) because of the mixture with Ricci tensor and
scalar curvature which emerge in the third line of the expression.

The next problem is to see what happens with the result (\ref{28})
on-shell. For this end we have to derive the classical equations of
motion and replace them into (\ref{28}). The equation of motion for
the theory (\ref{1}) with $\La=-3m^2$ have the form
\beq
R^{\mu\nu}-\frac{1}{2}Rg^{\mu \nu}=-3m^{2}g^{\mu \nu}
+\frac{1}{2}m^{2}g^{\mu
\nu}\left(\sqrt{g^{-1}f}\right)^{\al}{}_{\al}
-\frac{1}{2}m^{2}\left(\sqrt{g^{-1}f}\right)^{\nu \mu}.
\label{eqmotion}
\eeq
After using this relation in Eq.(\ref{28}), we
arrive at the following on-shell result
\beq
\nonumber
&&\overline{\Ga}^{(1)}_{div}\mid_{on\,\,shell} \,= \,-\,
\frac{\mu^{n-4}}{(4\pi)^{2}(n-4)} \int d^{n}x \sqrt{- g}\,
\Big[\frac{53}{45}E+m^{2}R^{(\la \mid\al\mid \si ) \be}f_{\si \tau}
{({\hat{A}}^{-1})}{}_{\la}{}^{\tau}{}_{(\al}{}^{\mid\rh\mid} f_{\be)
\rh}\Big]
\\
\nonumber
&-& \frac{m^{4}\mu^{n-4}}{8(4\pi)^{2}(n-4)} \int d^{n}x
\sqrt{-g}\Big[1.5 \cdot 32 \cdot 111 - 8f_{\al \la}
{({\hat{A}}^{-1})}{}^{\be \la , \al \si} f_{\be \si}
+ 0.4\cdot 77\big(\sqrt{g^{-1}f}\big)^{\mu}{}_{\mu}
\\
\nonumber
&+&
\frac{7}{5}\big(g^{-1}f\big)^{\mu}{}_{\mu}
+ 1.5\cdot 13 \cdot 29 \big(\sqrt{g^{-1}f}\big)^{\mu}{}_{\mu}
\big(\sqrt{g^{-1}f}\big)^{\nu}{}_{\nu}
\\ \nonumber
&+&
\big(g^{-1}f\big)^{\la}{}_{(\rh}
{({\hat{A}}^{-1})}{}_{\tau ) \la \mu}{}^{\si}
\big(g^{-1}f\big)^{\nu}{}_{\si}
\big(g^{-1}f\big)^{\rh}{}_{\th}
{({\hat{A}}^{-1})}{}^{\tau \th \mu}{}_{\al}
\big(g^{-1}f\big)^{\al}{}_{\nu}
\\
\nonumber
&+&
\big(g^{-1}f\big)^{\la}{}_{(\rh}
{({\hat{A}}^{-1})}{}_{\tau )\la}{}^{\mu}{}_{\si}
\big(g^{-1}f\big)^{\si}{}_{\nu}
\big(g^{-1}f\big)^{\rh}{}_{\th}
{({\hat{A}}^{-1})}{}^{\tau \th \nu}{}_{\al}
\big(g^{-1}f\big)^{\al}{}_{\mu}
\\
\nonumber
&-&
4\big(\sqrt{g^{-1}f}\big)^{\la}{}_{\la}\big(g^{-1}f\big)^{\be}{}_{\ph}
f_{\th (\al}(\hat{A}^{-1})_{\be )}{}^{\th \al \ph}
+\Big[\frac{3}{2}\big(\sqrt{g^{-1}f}\big)^{\la}{}_{\la}+16\Big]f_{\al \be} {({\hat{A}}^{-1})}{}^{\al \be ,\mu \nu}
f_{\mu \nu}
\\
\nonumber
&-&
4f_{\rh \th}\Big(\sqrt{g^{-1}f}\Big)^{\la}{}_{(\al}f_{\ph )\la}
{({\hat{A}}^{-1})}{}^{\rh \th \al \ph}
+2f_{\al \th}f_{\la \ph}\Big(\sqrt{g^{-1}f}\Big)^{\la}{}_{\rh}
{({\hat{A}}^{-1})}{}^{\rh \th \al \ph}
\\
\nonumber
&+&
2f_{\be \la}\big(g^{-1}f\big)^{\be}{}_{\ph}
\big(\sqrt{g^{-1}f}\big)^{\la}{}_{\th}
{({\hat{A}}^{-1})}{}_{\al}{}^{\th \al \ph}
+2\big(g^{-1}f\big)^{\th}{}_{\la}\big(g^{-1}f\big)^{\la}{}_{\ph}
\big(\sqrt{g^{-1}f}\big)^{\al}{}_{\rh}
{({\hat{A}}^{-1})}{}^{\rh}{}_{\th \al}{}^{\ph}
\\ \nonumber
&-&
2\big(g^{-1}f\big)^{\ta}{}_{\th}f_{\la \ph}
\big(\sqrt{g^{-1}f}\big)^{\al}{}_{\ta}
{({\hat{A}}^{-1})}{}^{\la \th}{}_{\al}{}^{\ph}
-\Big[\frac{1}{2}\big(\sqrt{g^{-1}f}\big)^{\mu}{}_{\mu}
+6\Big]\Big(g^{-1}f\Big)^{\al}{}_{\rh}
{({\hat{A}}^{-1})}{}^{\be \rh}{}_{(\al}{}^{\mid\ta \mid}f_{\be )\ta}
\\ \nonumber
&+&
2\big(\sqrt{g^{-1}f}\big)^{\al}{}_{\la}\Big(g^{-1}f\Big)^{\la}{}_{\rh}
{({\hat{A}}^{-1})}{}^{\be \rh}{}_{(\al}{}^{\mid \ta \mid}f_{\be )\ta}
+2\big(\sqrt{g^{-1}f}\big)^{\al}{}_{\la}\Big(g^{-1}f\Big)^{\be}{}_{\rh}
{({\hat{A}}^{-1})}{}^{\la \rh}{}_{(\al}{}^{\mid \ta \mid}f_{\be )\ta}
\\
&+& \Big[32-\frac{5}{2}\big(\sqrt{g^{-1}f}\big)^{\mu}{}_{\mu}\Big]
\Big(g^{-1}f\Big)^{\la}{}_{\be} {({\hat{A}}^{-1})}{}_{\al
\la}{}^{\al \si} \Big(g^{-1}f\Big)^{\be}{}_{\si}\Big].
\label{onshell}
\eeq
It is easy to see that the on-shell result does
not vanish as it was for the massless theory \cite{hove}. Moreover,
in the first line one can see the term which explicitly depends on
the Riemann tensor.

\section{Conclusion}
\label{con}
We have developed the background field method and
calculated the one-loop divergences for minimal massive gravity
models suggested in \cite{ghost free-2}. The divergences are
formulated in terms of geometrical invariants constructed from
metric and reference metric and contain the inverse matrix
(\ref{14-2}) which is an infinite power series in the reference
metric $f_{\mu\nu}$. There are no doubts that the divergences for
the non-minimal, more complicated actions of \cite{ghost free-2}
will have qualitatively the same structure. The final expression
(\ref{28}) shows that the UV completion of the massive gravity
theory would be essentially more complicated than the one of
Einstein quantum gravity. Along with the usual fourth-derivative
metric-dependent terms such completion should include also
dependence on the reference metric $f_{\mu\nu}$. Furthermore, this
field gains dynamics due to the mixture with curvature tensor
components. Therefore the counterterm (\ref{28}) can be considered
as the action functional defying dynamics of reference metric.

\section*{Acknowledgments}
I.B. is grateful to CAPES for supporting his visit to Juiz de Fora,
where the main part of this work has been done. Also he acknowledges
to RFBR grant, project No 12-02-00121, RFBR-Ukraine grant, project
No 11-02-90445 and grant for LRSS, project No 224.2012.2.
D.P. thanks FAPEMIG for supporting his PhD project. I.Sh. is
grateful to CNPq, CAPES and FAPEMIG for partial support.

\renewcommand{\baselinestretch}{0.9}


\begin {thebibliography}{99}

\bibitem{FP}
M. Fierz, Helv. Phys. Acta {\bf 12} (1939) 3;
\\
M. Fierz and W. Pauli, Proc. Roy. Soc. Lond. A {\bf 173} (1939) 211.

\bibitem{BD} D. G. Boulware and S. Deser,
Phys. Lett. {\bf B40} (1972) 227;
Phys. Rev. D {\bf 6} (1972).

\bibitem{ghost free-1}
C. de Rham, G. Gabadadze, Phys.Rev. D {\bf 82} (2010) 044020;

C. de Rham, G. Gabadadze, A.J. Tolley, Phys. Rev. Lett.
{\bf 106} (2011) 231101, arXiv:1011.1232[hep-th]; \
{\it Comments on (super)luminality.} arXiv:1107.0710[hep-th];
{\it Ghost free Massive Gravity in the St\'uckelberg language.} arXiv:1107.3820[hep-th];
{\it Helicity Decomposition of Ghost-free Massive Gravity} arXiv:1108.4521[hep-th].

\bibitem{ghost free-2}
S.F. Hassan, R.A. Rosen,
JHEP {\bf 1107} (2011) 009 arXiv:1103.6055[hep-th];
{\it Resolving the Ghost Problem in non-Linear Massive Gravity.}
arXiv:1106.3344[hep-th];
{\it Confirmation of the Secondary Constraint and Absence of Ghost
in Massive Gravity and Bimetric Gravity.} arXiv:1111.2070[hep-th];
\\
S.F. Hassan, R.A. Rosen, A. Schmidt-May,
{\it Ghost-free Massive Gravity with a General Reference Metric.}
arXiv:1109.3230[hep-th].

\bibitem{ghost free-3}
L. Alberte, A.H. Chamseddine and V. Mukhanov,
JHEP 1104 (2011) 004, arXiv:1011.0183 [hep-th];
\\
A.H. Chamseddine and V. Mukhanov,
JHEP 1108 (2011) 091; arXiv:1106.5868[hep-th].

\bibitem{ghost free-4}
M. Mirbabayi,
{\it A Proof Of Ghost Freedom In de Rham-Gabadadze-Tolley Massive Gravity},
arXiv:1112.1435[hep-th];
\\
A. Golovlev,
{\it On the Hamiltonian analysis of non-linear massive gravity.}
arXiv:1112.2134[gr-qc].

\bibitem{review} K. Hinterbichler,
{\it Theoretical Aspects of Massive Gravity.}
arXiv:1105.3735[hep-th].

\bibitem{DW-65}
B. S. DeWitt, {\it Dynamical theory of groups and fields}
(Gordon and Breach, New York, 1965).

\bibitem{article 9}
L. F. Abbott, Acta Phys. Pol. {\bf B13}, 33-50 (1982).

\bibitem{article 6}
I. L. Buchbinder, S. D. Odintsov and I. L. Shapiro,
{\it Effective action in quantum gravity}
(IOP Publishing , Bristol, 1992).

\bibitem{hove}
G. `t Hooft and M. J. G. Veltman, Annales Poincare Phys. Theor. A
{\bf 20}, 69 (1974).

\bibitem{BV}
A.O. Barvinsky, G.A. Vilkovisky, Phys.Repts. {\bf 119}, 1 (1985).





\end{thebibliography}
\end{document}